\documentclass[reprint]{revtex4-1}
\usepackage{graphicx} 
\usepackage{amssymb}
\usepackage{wasysym}
\newcommand{\chem}[1]{\ensuremath{\mathrm{#1}}}

\begin{document}
\title{
Direct and alignment-insensitive measurement of cantilever curvature} 
\author{Rodolfo I. Hermans}
\email[]{r.hermans@ucl.ac.uk}
\affiliation{London Centre for Nanotechnology, University College London}
\affiliation{Department of Physics and Astronomy, University College London}

\author{Joe M. Bailey}
\affiliation{Centre for Mathematics and Physics in the Life Sciences and Experimental Biology, University College London}
\affiliation{London Centre for Nanotechnology, University College London}

\author{Gabriel Aeppli}
\email[]{gabriel.aeppli@ucl.ac.uk}
\affiliation{London Centre for Nanotechnology, University College London}
\affiliation{Department of Physics and Astronomy, University College London}


\date{\today}

\begin{abstract}
We analytically derive and experimentally demonstrate a method for the simultaneous measurement of deflection for large arrays of cantilevers. The Fresnel diffraction patterns of a cantilever independently reveals tilt, curvature, cubic and higher order bending of the cantilever. It provides a calibrated absolute measurement of the polynomial coefficients describing the cantilever shape, without careful alignment and could be applied to several cantilevers simultaneously with no added complexity. We show that the method is easily implemented, works in both liquid mediums and in air, for a broad range of displacements and is especially suited to the requirements for multi-marker biosensors.
\end{abstract}

\maketitle 

Silicon-based microfabrication has enabled not only the electronics revolution but also made micromechanics nearly as ubiquitous, with applications from motion sensing to biochemical analysis\cite{Lavrik2004,Raiteri2001}. The atomic force microscope \cite{Binnig1986,Hwang2009}, where the motion of a small tip at the end of a cantilever traces nanoscale features on surfaces, provides the fundamental paradigms for nanomechanical metrology, including optical readouts of cantilever displacement and bending. Such readout typically requires careful and costly alignment of mechanical and optical elements. Here we demonstrate an alternative approach based on the recognition that in many applications we are interested not so much in the motion of a tip as in the overall curvature of a cantilever. In particular, near field imaging of entire cantilevers yields diffraction patterns providing precise measures of the tilt, curvature and higher order bending components of cantilevers. Even while we have used very inexpensive components and no careful alignment is required, we obtain sensitivity to nm-scale motion of the cantilever end.

There are many methods to measure the displacement of microstructures such as AFM cantilevers\cite{Hwang2009}. They are based on various physical principles, including optics\cite{Li2009}, piezoresistance\cite{Linnemann1995,Rasmussen2003}, field-effect transistors\cite{Tark2009} and capacitance\cite{Blanc1996}. The method most commonly used, and still the most sensitive and reliable, is the optical lever or optical beam deflection technique (OBDT)\cite{Putman1992,Beaulieu2007} --implemented in the AFM market-- and optical interferometry \cite{Yaralioglu1998a,Helm2005b,Hoogenboom2005}. No detection method is optimal for all types of measurements and the growing use of cantilevers as multiplexed biosensors imposes its own challenges\cite{Ndieyira2008c}, particularly in the case of arrays of several cantilevers. Cantilever arrays are used to obtain simultaneous detection of different targets, increase statistical significance, in-situ control and for differential measurements\cite{McKendry2002,ZhangJ.2006}. They require detection systems with a complexity that typically scales with the number of cantilevers in the array. Examples of such systems include arrays of illuminating lasers with a single multiplexed detector\cite{Lang1998} or 2D scanners of a single laser beam employing voice-coil actuators\cite{Martinez2010}, which are simply extensions of OBDT for cantilever arrays.

The main drawback of the common optical techniques is the difficulty of accurately aligning each illuminating source with its corresponding cantilever and detector, making it difficult to measure large numbers of cantilevers without an elaborate pre-measurement protocol. A further limitation, particularly of OBDT (Fig~1a), comes from the fact that the observed quantity is the local change in angle and displacement of the lever at the point of illumination, and therefore cantilever tilt and bending cannot be distinguished in a single measurement. Also, a bending model must be assumed to estimate the true beam curvature. Efforts to measure the whole cantilever profile have partially addressed the latter issue using scanning or an array of Light-Emitting Diodes (LED)\cite{Jeon2004b}. Research has revealed that commonly assumed bending profiles may not be as realistic as expected\cite{Mertens2005}, providing further justification for the development of a simple technique for imaging cantilever bending.

\begin{figure}
\includegraphics[width=\columnwidth]{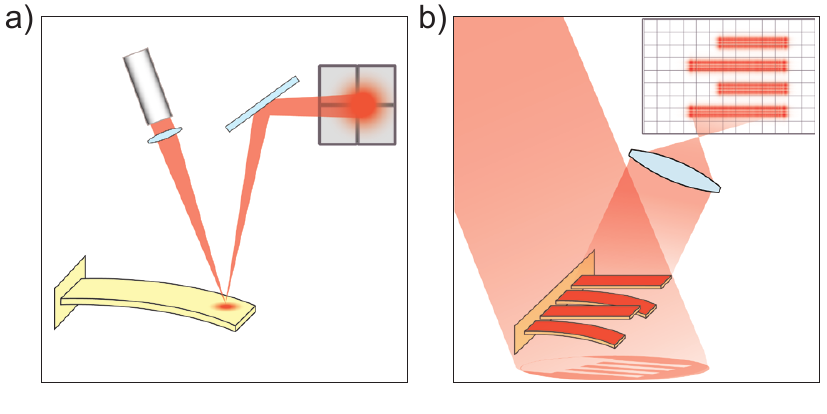}
\caption{Concepts for standard and proposed optical detection methods. a) The optical beam deflection technique (OBDT) requires a focused beam carefully aligned to a cantilever and the center segmented photo-diode. b) In the \emph{No-Alignment Nearfield Optical Bending Estimation} (NANOBE) technique, a broad beam illuminates the whole cantilever array generating a diffraction pattern projected onto a CMOS detector. The intensity profile of the diffraction pattern is insensitive to misalignments but sensitive to the details of the cantilever bending.}\label{figN1}
\end{figure}

In this Letter we describe a method for the direct and calibrated measurement of cantilever bending profiles that requires no scanning or alignment of the illumination and does not assume particular bending models. We show that an out-of-focus image of the whole cantilever array is enough to estimate at least the first four coefficients of a polynomial describing the cantilever shape for each of the cantilevers in an array independently, making it suitable for multi-marker biological essays. We name this technique NANOBE for \emph{No-Alignment Nearfield Optical Bending Estimation}.


To detect bending profiles we illuminate a cantilever array with a homogeneous monochromatic plane wave and measure the intensity pattern of the reflected light (Fig~1b). Based on the Huygens-Fresnel principle, we model the reflected component from the finite-size cantilever as a rectangular source in the plane $(\xi, \eta)$. We express the wave amplitude in the observing plane $(x,y)$ as the convolution integral~\cite{goodman2005introduction}

\begin{equation}\label{eqn:fresnel-conv}
    U(x, y) = \frac{e^{i k z}}{i \lambda z} \iint\limits_{-\infty}^{\infty}  U(\xi, \eta)\, e^{ i \frac{\pi}{\lambda z} \left[ (x-\xi)^2+(y-\eta)^2 \right]} d\xi d\eta.
\end{equation}

where $U(\xi, \eta)$ is the function defining the amplitude and phase at the source. We model the shape of the cantilever as a rectangle of dimensions $(w,l)$ and the bending profile by a polynomial function $P(\xi)=\sum c_i {\xi}^{i}$ with $i\geq1$  ;  although here we assume bending only along the $\xi$ direction which is the long axis of the cantilever, our results can be generalized to $P(\xi, \eta)$ taking account of arbitrary curvature. The bending of the surface introduces a difference in the optical path, which introduces a phase $\phi = 4\pi\lambda^{-1} P(\xi)$. The field amplitude at the source is well described by

\begin{equation}
    U(\xi, \eta) = \text{rect}\left(\frac{\xi}{l} \right)e^{\frac{4 \pi i}{\lambda} P(\xi) }\,\, \text{rect}\left(\frac{\eta}{w} \right)
\end{equation}

\begin{figure}
\includegraphics[width=\columnwidth]{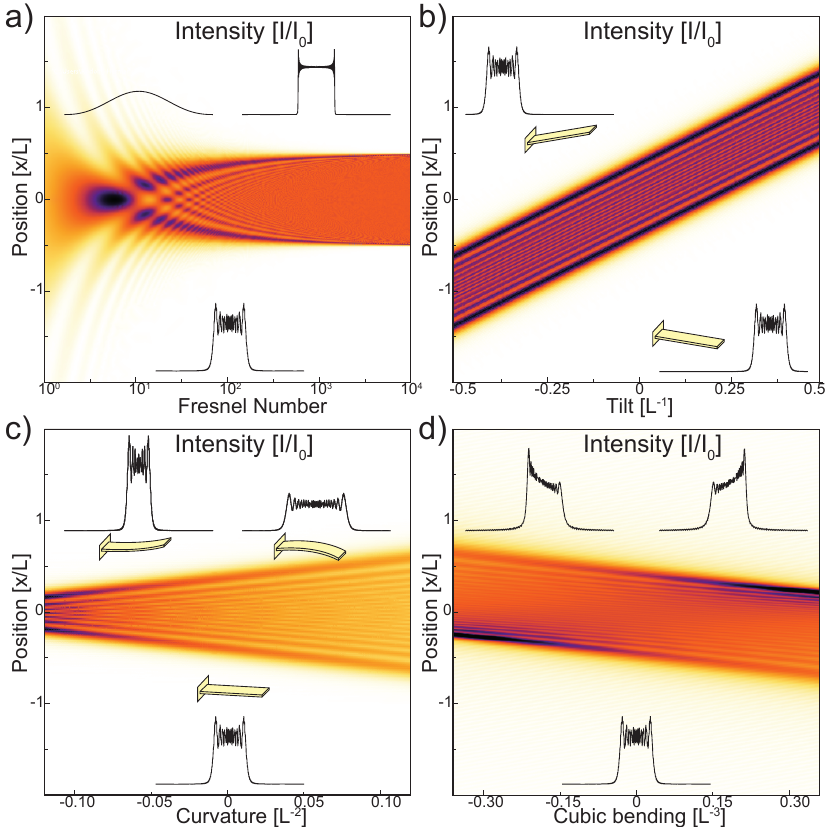}
\caption{Intensity map $I(x)$, where x is pixel position along cantilever direction on the detector, for a cantilever as a function of different relevant variables. Insets are vertical cross-sections of the intensity map showing the intensity profiles for specific abscissa values. The horizontal axes are: a) Fresnel number $F=L^2 (\lambda z)^{-1}$, where $L$ is the cantilever length, and $z$ the distance from the cantilever at which the intensity is measured, b) $c_1$ the cantilever tilt, c) $c_2$ the cantilever curvature and d)  $c_3$ the cubic bending. }\label{figN2}
\end{figure}

where $ \text{rect}(x)=1$ for $|x|<1/2$ and zero otherwise.
We concentrate attention on the $x$ coordinate along the cantilever and rewrite equation \ref{eqn:fresnel-conv} to separate variables $U(x,y) = -i \exp(i k z)\, \mathcal{I}(x) \mathcal{I}(y)$ and considering that the observed intensity $I(x,y)=\left|U(x,y)\right|^2 = \left|\mathcal{I}(x)\right|^2 \,\left|\mathcal{I}(y)\right|^2$ we analyze the value of

\begin{equation}\label{eqn:fresnel-wx}
\mathcal{I}(x) = \frac{1}{\sqrt{\lambda z}} \int\limits_{-L/2}^{L/2} \exp\left[ i \frac{\pi}{\lambda z} (x-\xi)^2 + i\frac{4 \pi}{\lambda} P(\xi) \right]\, d\xi
\end{equation}

We now use the first two terms of  $P(\xi)$ to complete a squared binomial for $\xi$ and for convenience rewrite equation~\ref{eqn:fresnel-wx} as

\begin{equation}\label{eq:AmplFresnelPattern}
\mathcal{I}(x) = A  \int\limits_{-L/2}^{L/2} e^{ i \frac{\pi}{\lambda z} \left( m \left(\frac{x-s}{m}-\xi \right)^2  + 4 z \sum\limits_{i\geq3} c_i {\xi}^{i} \right)} d\xi .
\end{equation}

with $m = 1+4 c_2 z$ , $s=2 c_1 z$ and

\begin{equation}
 A =  (\lambda z)^{-\frac{1}{2}}  \exp\left[ i \frac{4 \pi}{\lambda \,   m}  \left(c_2 x^2 + c_1 x - c_1^2 z \right) \right].
\end{equation}

If we neglect cubic and higher order terms in the polynomial description of the cantilever curvature \emph{i.e.} assuming $c_i\approx 0 \,\,\forall \, i>2$, we can introduce a change of variables and a corresponding change in integration limits to find a familiar result. Using

\begin{equation}\label{eq:ChangeVariables}
\alpha(\xi) = \sqrt{\frac{2m}{\lambda z}} \left(\frac{x-s}{m}-\xi \right)
\end{equation}

we obtain the solution for a rectangular slit

\begin{equation}
I(x) = \frac{\left( (C(\alpha_2)-C(\alpha_1))^2 + (S(\alpha_2)-S(\alpha_1))^2 \right)}{2 m},
\end{equation}

where $C(\alpha)$ and $S(\alpha)$ are \emph{Fresnel Integrals} defined as $C(\alpha_i)=\int_0^{\alpha_i } \cos \left(\pi  \alpha ^2/2\right) d\alpha,$ and $S(\alpha_i)=\int_0^{\alpha_i } \sin \left(\pi  \alpha ^2/2\right) d\alpha$ and the integration limits $\alpha_1 = \alpha(-L/2)$ and $\alpha_2 = \alpha(L/2)$

\begin{eqnarray}
  \alpha_1  &=&\sqrt{2} \sqrt{\frac{4 c_2 z+1}{\lambda  z}} \left(-\frac{L}{2}-\frac{x-2 c_1 z}{4 c_2 z+1}\right) \\
  \alpha_2  &=&\sqrt{2} \sqrt{\frac{4 c_2 z+1}{\lambda  z}} \left(\frac{L}{2}-\frac{x-2 c_1 z}{4 c_2 z+1}\right)
\end{eqnarray}
The inset of figure~\ref{figN1}b shows calculated 2D patterns given by $I(x,y)= \left|\mathcal{I}(x)\right|^2 \,\left|\mathcal{I}(y)\right|^2$ resembling experimentally observed patterns.
Figure~\ref{figN2}a shows $I(x)$ for a rectangular slit or a flat cantilever \emph{i.e.} $c_i=0 \,\, \forall i$ as a function of the Fresnel number $F=L^2 (\lambda z)^{-1}$, where $L$ is the cantilever length, and $z$ the distance from the cantilever at which the intensity is measured (insets are intensity profiles for specific abscissa value). In what follows we analyze the influence of each coefficient $c_i\neq0$ on the observed diffraction pattern.

From the equation~\ref{eq:ChangeVariables} itself and the figure~\ref{figN2}b we see that a non vanishing tilt $c_1$ results in a shift $s=2 c_1 z$ of the diffraction pattern, as expected from the tilt of any reflecting surface. Similarly from figure~\ref{figN2}c, provided $z\neq0$, a non-zero $c_2$ causes a magnification $m = 1+4 c_2 z$ in the pattern size and $m^{-1}$ in the intensity. The magnification does not come from the negligible displacement of the cantilever ends, but is intrinsic to the curvature of the reflecting surface. Both effects do not otherwise distort the shape of the diffraction pattern.

Higher order contributions to the curvature are more easily studied by numerically integrating equation~\ref{eq:AmplFresnelPattern}. Figure~\ref{figN2}d shows the Fresnel diffraction pattern for cantilever with a deflection profile $P(x) \propto x^3$. Cubic bending shown in figure~\ref{figN2}d features intensity gradients and pattern shifts, both proportional to $c_3$. Quartic bending (not shown) creates a curvature in the intensity profile as well as a change in the pattern length.


\begin{figure}
\includegraphics[width=\columnwidth]{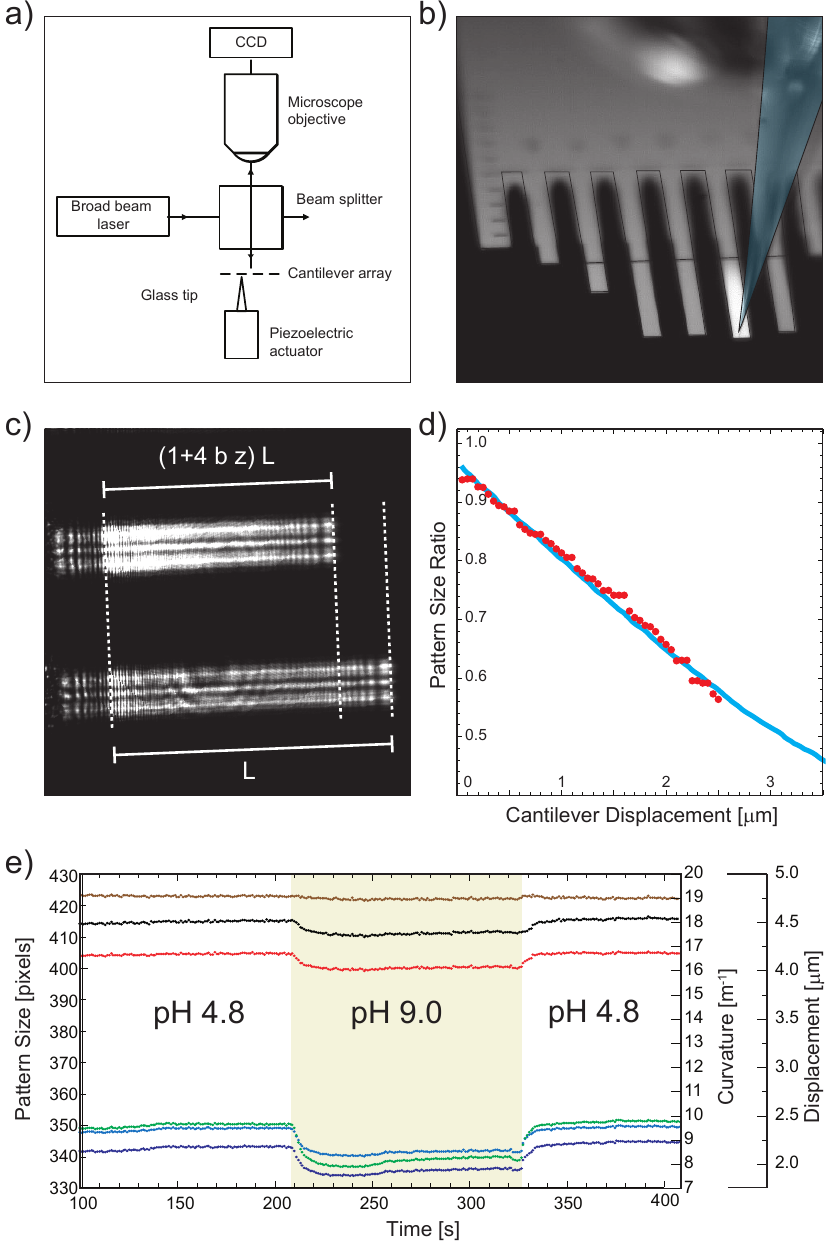}
\caption{Experimental measurements using NANOBE method. a) Setup schematics: A cantilever array is illuminated with a broad laser beam through a cubic beam-splitter and the reflected light captured by a lens and CCD. b) A single cantilever is pushed with a glass tip mounted over a calibrated piezoelectric actuator. c) Diffraction images ($640\times528$ px) of the pushed cantilever (top pattern) are shorter than the patterns from the relaxed cantilever by a factor of $(1+4 b z)$. This corresponds to a lensing effect from the curved mirror formed by the cantilever. d) The experimentally measured pattern size changes linearly with the cantilever tip displacement closely overlapping the model. e) The curvatures of six cantilevers as a function of time. Cantilevers functionalized with MHA (top) and HDT (bottom) have different static bending and different sensitivity to pH changes.}\label{figN3}
\end{figure}

\begin{figure}[tbh]
\includegraphics[width=\columnwidth]{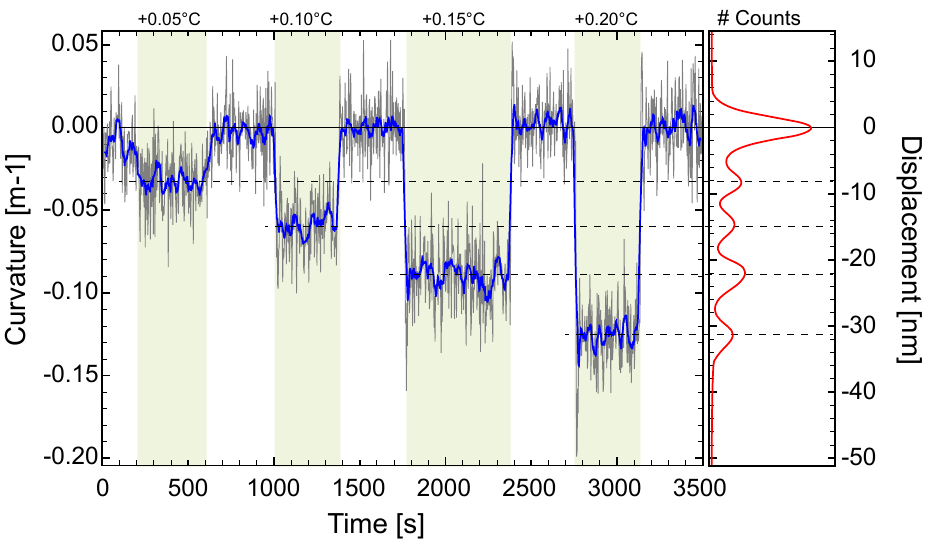}
\caption{The cantilever bending caused by small temperature changes are easily resolvable. The temperature protocol was set to $25^{\circ}C$,$25.05^{\circ}C$,$25.10^{\circ}C$,$25.15^{\circ}C$ and $25.20^{\circ}C$. The light trace corresponds to the raw data and the darker trace is a 10 point moving average. The right hand side graph shows a kernel density estimation of raw data demonstrating that the measured curvatures feature perfectly distinguishable distributions."}\label{figN5}
\end{figure}


Figure~3a illustrates the experiment where a broad ($16.8 mm$) collimated laser beam ($\lambda=660 nm$) illuminates a cantilever array. A 4X microscope objective and a CMOS sensor focused in a plane at a distance $z$ from the cantilever maps the diffraction pattern from the cantilevers. The array (purchased from IBM) consisted of up to eight rectangular silicon (100) cantilevers, each measuring $500\mu m$ in length, $100 \mu m$ in width, and $0.9 \mu m$ in thickness and a nominal spring constant of $0.02 N/m$. The cantilevers were coated on one side with a $2 nm$ titanium adhesion layer and then $20 nm$ of gold. The diffraction pattern size is calculated in units of pixels by software calculating the distance between intensity threshold crossings or, more preferable, using the second central moment of the intensity profile $I(j)$~\cite{NANOBE-SupplMat}.

\begin{equation}
S_o=    2 \sqrt{3\frac{\sum I(j) (j-\mu)^2}{\sum I(j)}}
\end{equation}

We test the method experimentally for deflections in excess of one micron by measuring the diffraction pattern as a function of cantilever bending where we poke the cantilever end in air with a glass tip (Fig~3b) attached to a well-calibrated piezoelectric actuator (P-363 PicoCube XYZ Piezo Scanner by Physik Instrumente GmbH). The concentrated load at the free end is expected to cause a bending profile proportional to $x^2(3L-x)$, where $L$ is the cantilever length\cite{Green1996}. Along with the controlled displacement we observe both a change of the size of the diffraction pattern as well as a gradient in the intensity, characteristic of both parabolic and cubic bending (Fig~3c top pattern). The change in pattern size displays the approximately linear predicted relationship with the displacement caused by the actuator. Figure~3d shows the overlap of experimental data (dots) and numerically integrated model (line).

We further test the method in fluid by functionalizing the surface of individual cantilevers with either mercaptohexadecanoic acid \chem{HS(CH_2)_{15} COOH}, here abbreviated MHA, or hexadecanethiol \chem{HS(CH_2)_{15} CH_3}, here abbreviated HDT, via incubation in an array of glass microcapillaries. The protonation/deprotonation of MHA at different pH causes a differential surface stress inducing cantilever bending. HDT has identical chain length and similar packing density to MHA but differs in the terminal methyl group which is non-ionizable and therefore is used as reference~\cite{Watari2010}. The cantilever array is then mounted in a liquid cell with a sapphire window to allow illumination and imaging. An automated system of syringes and valves was developed to control the delivery of sodium phosphate mono and dibasic solution pH 4.8 and pH 9.0 at a constant rate of $43 \mu L/\text{min}$ and controlled temperature of $25.00\pm0.01^{\circ}C$. The supplementary materials~\cite{NANOBE-SupplMat} show a schematic as well as photograph of the instrumental prototype which we have constructed for these measurements. Figure~3e shows the distinctive changes in pattern size as we alternately flow solutions with pH 4.8 and pH 9.0. The main changes in curvature are attributed to the differential stress caused by protonation/deprotonation. The mean relative deflection for MHA cantilevers is calculated to be $249.8 \pm 8.6\,nm$, consistent with previous results~\cite{Fritz2000}. The mean relative deflection for HDT is somehow smaller than previous results ($137.2 \pm 4.8\,nm$) giving a differential stress between MHA and HDT-coated cantilevers of around $22.0 \pm 1.9\,mN/m$, 52\% bigger than expected~\cite{Watari2007}. Observing the initial curvatures in figure 3e, we see that NANOBE reveals that HDT and MHA functionalization causes different static bending of the cantilevers as well as different sensitivities to pH. The variation in static and dynamic bending among cantilevers with the same functionalizations is commonly observed and reveals the difficulties in preparing consistent surface modification in small areas. The inconsistency upon functionalization of nominally identical cantilevers remains an open question of utmost  importance but is beyond the scope of this work. Our implementation of NANOBE allows direct observation of unexpected objects on the surface, lateral torsion and some but not all other sources of error typically undetectable on OBDT.

To demonstrate the resolution of small deflections in liquid we step the temperature of the cantilever in increments of $0.05\pm0.01^{\circ}C$ around $25^{\circ}C$. The different thermal expansion of silicon and gold causes the micro-structure to bend. Figure~4 shows that the measured deflection is well resolved even for these small changes in temperature. The inset on the right hand side shows the density of measured deflections, featuring well-resolved maxima and negligible overlap of the individual distributions. The cantilever is found to bend around $150 nm/^{\circ}C$, in good agreement with previous measurements and models~\cite{Ansari2011}.


The main output of the NANOBE method is the size of a diffraction pattern which is independent of the region of the detector that is used to capture it, and therefore insensitive to small misalignments of the light source or detector and independent of the orientation of the cantilever. The pattern size measured in units of pixels is directly translated to an absolute value for the curvature using

\begin{equation}
   c_2= (S_o-S_c)/(4 z S_c)
\end{equation}

where $S_o$ is the observed pattern size in pixels, $S_c=\rho L_c$, where $L_c$ is the cantilever length  and $\rho$ resolution of the sensor in pixels per length, and $z$ is the distance to focus. In contrast to the standard laser beam deflection method where the optical lever arm needs to be carefully fixed and measured, our method is entirely self-calibrated, i.e. the known geometry of the cantilever array gives the conversion ratio between length and pixels, and the cross section of the diffraction pattern reveals the effective value of distance $z$.  Nonetheless, if there are any doubts, these values are also available to the experimentalist by physically measuring the geometry of the experimental setup.
The changes of the tilt $c_1$ are also available from the centroid of the diffraction pattern, and the higher orders of curvature $c_3$ and above, can be estimated from the overall intensity gradients for each pattern.

Provided the Fresnel number is large enough (Fig~2a), the diffraction patterns will be localized even for narrow cantilevers ($\sim10\mu m$) with negligible cross talking between adjacent cantilevers, therefore allowing independent analysis of simultaneously imaged patterns.

The resolving power of the system is proportional to the magnitude of $z$, to the number of pixels covered by the pattern and inversely to the noise level of the detector. Our experiments were performed with a simple USB $1280\times1024$ px Monochrome CMOS (Thorlabs DCC1545M) at 1 frame per second where each pattern covered only around 400 pixels; therefore, a much improved resolution can be expected from a high-end detection system and higher sampling rates.


We have invented a near-field method for measuring cantilever bending which is much more robust and simpler to implement than the standard optical beam deflection methods. The method is model-free and gives independent values for tilt, curvature and higher order bending. It relies on curved mirror diffraction and is insensitive to misalignment, opening the way to a variety of devices for nanometrology, inexpensive from both the manufacturing and operational points of view. We envision future devices ranging from force microscopes to biochemical assays where cantilevers of various geometries that could even be weakly tethered (or untethered) to substrates using inexpensive hardware comparable to a portable CD player and a computer webcam.

\begin{acknowledgments}
G.A. and R.H. thank support from grant EPSRC EP/G062064/1, ``Multi-marker Nanosensors for HIV'', May 2009.
R.H and G.A on 28 March 2012 filed a patent application (``Measurement of micro-structure curvature'' 12054921.2) for the methods here described.
All authors thank Dr. Joseph Ndieyira and Dr. Samadhan Patil for metal-coating some of the cantilever arrays used in this work.
\end{acknowledgments}


%

\end{document}